\documentclass[times,twocolumn,final,authoryear]{elsarticle}
\usepackage{lineno}
\usepackage{soul}
\usepackage{sistyle}
\usepackage{amsmath}
\usepackage{jasr}
\usepackage{framed,multirow}
\usepackage{amssymb}
\usepackage{latexsym}

\usepackage{url}
\usepackage{xcolor}
\definecolor{newcolor}{rgb}{.8,.349,.1}

\usepackage{hyperref}

\journal{Advances in Space Research}

\begin{document}
\newcommand{\Cnm}{\overline{C}_{nm}}
\newcommand{\Snm}{\overline{S}_{nm}}
\newcommand{\Pnm}{\overline{P}_{nm}}
\newcommand{\Dsumz}{\sum_{n=0}^{\infty}\sum_{m=0}^n}
\newcommand{\Dsumo}{\sum_{n=1}^{\infty}\sum_{m=0}^nz}
\verso{Robert Spero}
\begin{frontmatter}
\title{Point-mass sensitivity of gravimetric satellites}
\author{Robert Spero}
\address{robert.spero@jpl.nasa.gov \\ Jet Propulsion Laboratory, Pasadena, CA 91109}
\received{12 December 2020}


\begin{abstract}
Frequency-domain expressions are found for gradiometer and satellite-to-satellite tracking measurements of a point source on the surface of the Earth.    The maximum signal-to-noise ratio as a function  of noise in the measurement apparatus is computed, and from that the minimum detectable point mass is inferred. 
 A point mass of magnitude $M_3=\SI{100}{Gt}$ gives a signal-to-noise ratio of 3 when  a GOCE-like gradiometer passes directly over the mass.  
 On the satellite-to-satellite tracking mission GRACE-FO $M_3=\SI{1.3}{Gt}$ for the microwave instrument and $M_3= \SI{0.5}{Gt}$  for the laser ranging interferometer.
The sensitivity of  future GRACE-like missions with different orbital parameters and improved accelerometer sensitivity is explored, and the optimum spacecraft separation for detecting point-like sources is found.  The  future-mission benefit of improving the accelerometer sensitivity for measurement of non-gravitational disturbances is shown by the resulting reduction of $M_3,$ to as small as \SI{7}{Mt} for \SI{500}{km} orbital altitude and optimized satellite separation of \SI{900}{km}.
   \end{abstract}
\end{frontmatter}

\section{Introduction}

A global gravity map is the principal data product of  satellite gravity missions.    Previously  CHAMP (\cite{reigber2003champ}), GRACE (\cite{tapley503}), and GOCE (\cite{drinkwater2006goce})  collected data to map the Earth's gravity, and GRAIL (\cite{konopliv2013jpl}) measured the Moon's gravity.  Currently GRACE Follow-On (GRACE-FO, ~\cite{landerer2020GFO})  is  extending the GRACE data record, with increased ranging precision afforded by its laser ranging interferometer (LRI, ~\cite{abich2019orbit}).  

As pointed out by \cite{watkins2015improved}, the most commonly used  method of analyzing satellite gravity data is based on global gravity fields expressed in terms of spherical harmonic basis functions.  An alternative to spherical harmonics is the mass concentration, or mascon, model.   Starting with  \cite{wong1971surface}, the mascon approach was applied to single-satellite lunar orbital measurements to infer the surface gravity of the moon.   Mascons can be modeled as many discrete sources (\cite{pollack1973spherical}, \cite{watkins2005grace})  that cover the globe, or used to solve for regional fields.  \cite{han2013determination} applied mascons to GRAIL data to solve for  regional fields of the Moon.

Though they differ slightly in assumptions and results, the spherical harmonic and mascon methods are constructed to answer the same question: what is the  gravity field that is most consistent with measurements?   Here we address a different question: what is the limit to measurement precision of a  point-like mass on the surface?  This is an  artificial model,  a single mascon, that is not directly applicable  to the geodetic agenda of measuring the Earth's gravity.  We do not attempt to replicate the global gravity-field inversion achieved by the usual  many-mascon analysis.  Rather, the motivation for this analysis is twofold:  to provide a single-number figure of merit, namely the minimum detectable isolated point mass perturbation, and to find the optimal filter for such a detection.
 The minimum detectable mass $M_3$ is defined as the point mass that a gives a signal-to-noise ratio $\rho= 3$ in a single orbital pass directly over the point mass.  It is calculated by applying the Wiener optimal filter to the problem of detecting a signal of known waveform, against a background specified by instrument noise power spectral density (\cite{wainstein1970extraction}). A comparison of $M_3$ for different orbital configurations and instrument sensitivities guides the design of future missions.  Additionally, we find the optimum  satellite separation in GRACE-like missions for a specified instrument noise power spectral density.

\section{Gradiometer Mass Sensitivity}
\label{grad-sec}
\begin{figure}[!ht]
 \center
\includegraphics[width=0.4\textwidth]{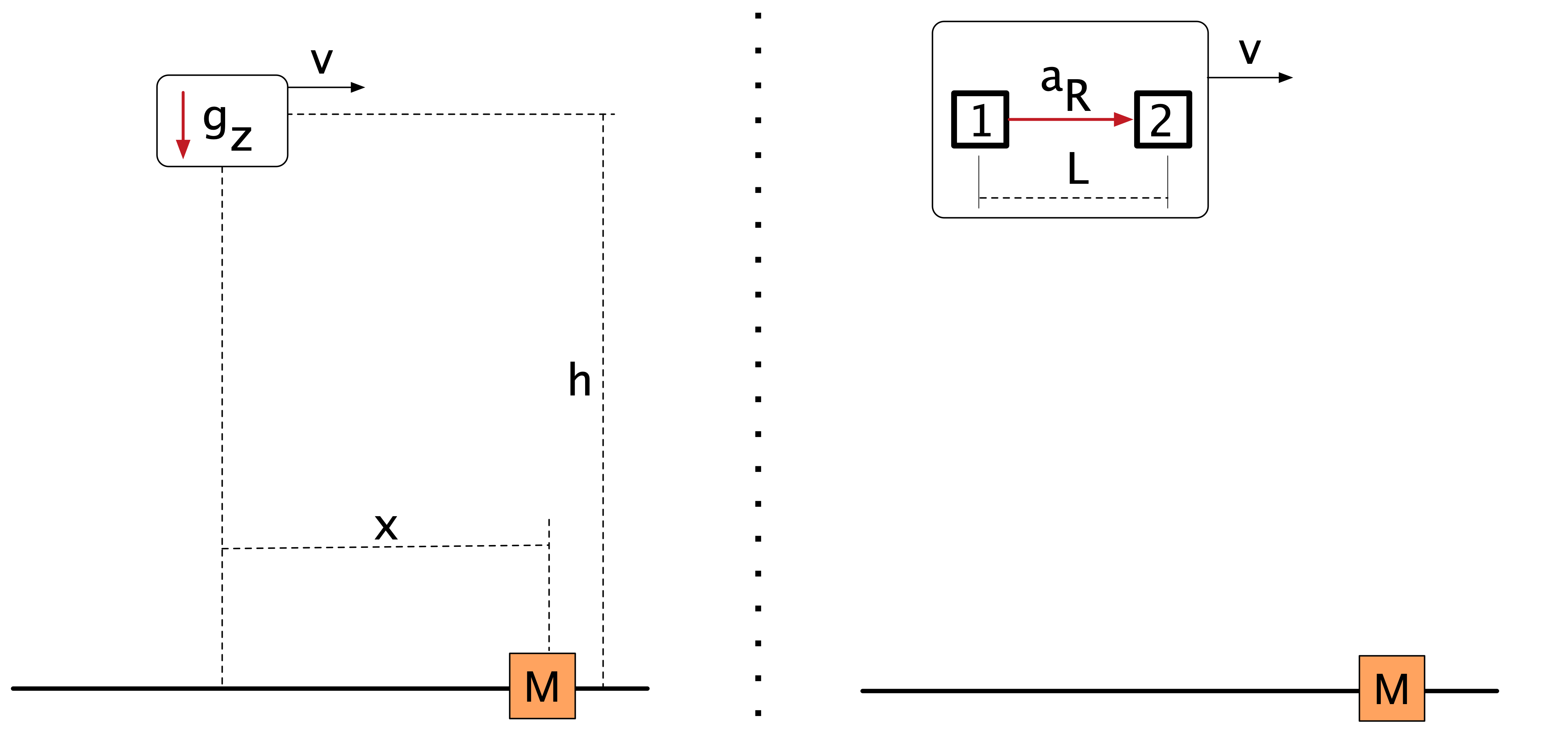}
\caption{Measurement of gravitational field from a point mass $M$ at along-track distance  $x$, altitude $h,$ and velocity $v.$  {\em Left:} Vertical gradient $g_z.$  {\em Right:} Differential acceleration  $a_R$ between spacecraft 1 and 2 with average separation $L$.   \label{sig-parameters}}
\end{figure}

Consider a  gradiometer flying directly over a point mass $M$ at altitude $h,$   Figure~\ref{sig-parameters}  left.
At orbital altitude $h=\SI{330}{km},$ the along-track velocity $v$ is the orbital velocity $v_o = \SI{7.7}{km/s}$ and the along-track  distance $x$ changes at approximately constant rate, $x=v_ot.$  The acceleration at the spacecraft in the vertical, $z$ direction is $a_z=-GMz/(x^2+z^2)^{3/2},$   and the gradient in the $z$ direction is $g_z=da_z/dz =GM[ 3z^2(x^2+z^2)^{-5/2}-(x^2+z^2)^{-3/2}].$ $G=$ Newton's constant of gravitation. Substituting $z\rightarrow h$ and $x\rightarrow v_ot$,
\begin{equation}
\label{gradient-time}
g_z(t)=\kappa_g M\left[
3(1+[f_ht]^2)^{-5/2}
-(1+[f_ht]^2)^{-3/2}
\right]
 \end{equation}
 where $\kappa_gM=GM/h^3= \SI{1.85e-3}{mE}$ with $M=\SI{1}{Gt}$ and   $f_h=v_o/h=\SI{23}{mHz};\ \SI{1}{mE}=\SI{1e-12}{s^{-2}}.$   At $t=0$, when the gradiometer is directly above the source mass, the gradient reaches its maximum value $g_z=2\kappa_gM.$  At $x=\pm\sqrt{2}h,\ g_z=0.$
 
The Fourier transform of $p(t)$ is defined by ${\mathcal F}[p(t)]\equiv p(f)=\int_{-\infty}^{\infty}dt\,p(t)\exp(-2\pi i f t).$  Applying the Fourier transform to  $g_z(t)$:
\begin{equation}
\label{FT-gz}
g_z(f)=\kappa_gM\frac{4\pi f }{f_h^2}\left[
 K_1\left(\frac{2\pi f} {f_h}\right)
 +\frac{2\pi f}{f_h}K_0\left(\frac{2\pi f}{f_h}\right)
 \right]
\end{equation}
where  $K_n$ is the modified Bessel function of the second kind, order $n$. 
From~\cite{abramowitz+stegun}, Section~9.7.2, 
$K_n(z)\sim\sqrt{\pi/(2z)}e^{-z},$ where $\sim$ indicates approximately equal for large $z.$ At   frequency $f\gg f_h,$ the measurement response is attenuated approximately exponentially with $e$-folding frequency $f_h/(2\pi) = \SI{3.7}{mHz}.$ This corresponds to harmonic order $N = f_h/(2\pi f_1) =20,$
 where $f_1=\SI{0.183}{mHz}$ is the orbital frequency.

\begin{figure}[!ht]
\center
     \includegraphics[width=0.5\textwidth]{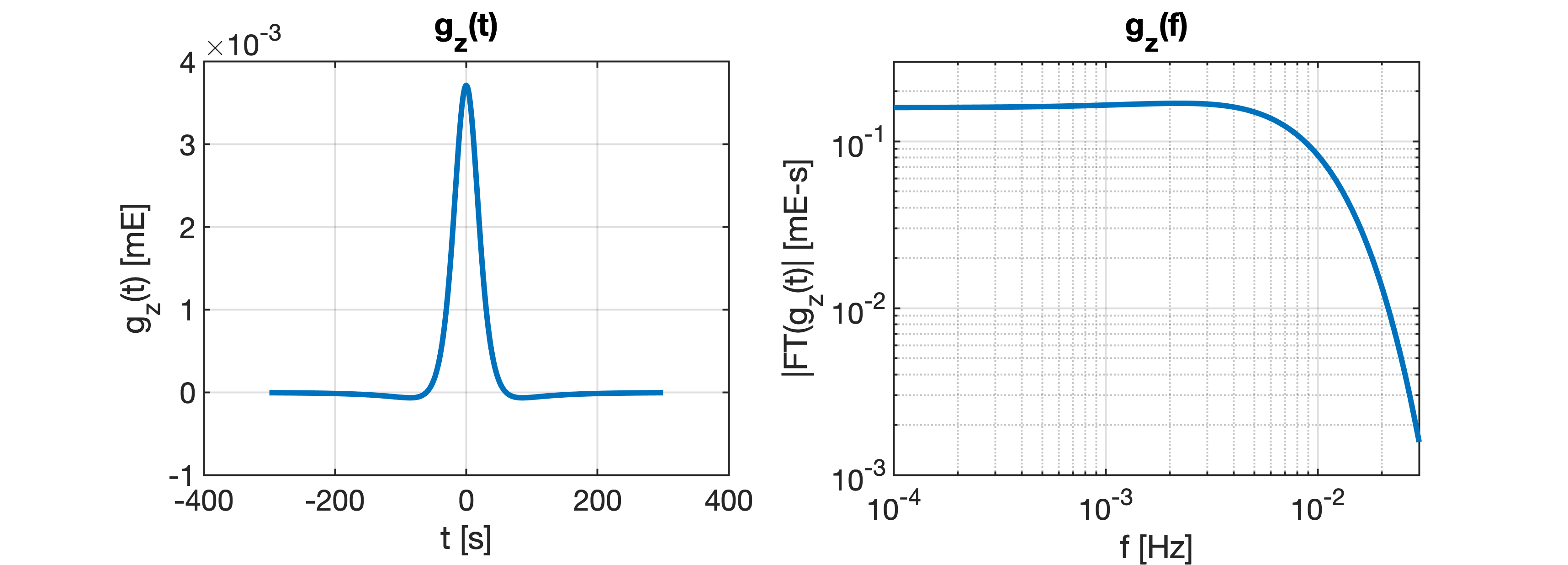}
     \includegraphics[width=0.5\textwidth]{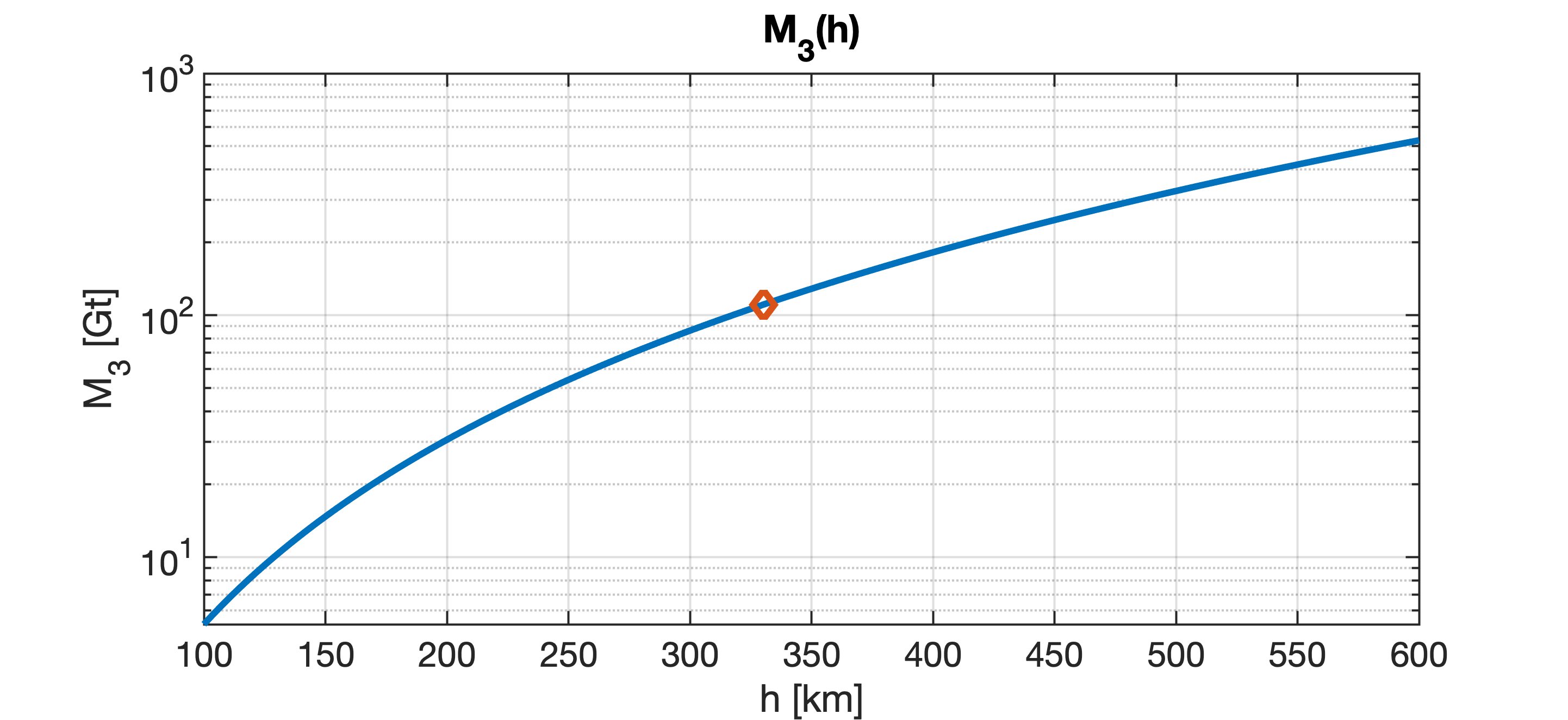}
     \caption{Orbiting gradiometer vertical gradient response to a point mass. {\em Upper-left}: $g_z(t),\ z$-direction gradient in time domain, Equation~\ref{gradient-time}; {\em Upper-right}:  $g_z(f)= $ Fourier transform, Equation~\ref{FT-gz};  {\em Lower}:  $M_3= $ Detectable mass with signal-to-noise = 3, Equation~\ref{M3}.\label{gradiometer-response}}
\end{figure}

The signal-to-noise ratio $\rho$ depends on the signal and the power spectral density of the gradiometer noise, $S_g(f).$  Define the frequency-dependent signal-to-noise-ratio density SNRD as
\begin{equation}
\label{Wg-def}
W_{g_z}(f)=\frac{|g_z(f)|^2}{S_g(f)}.
\end{equation}

In general,  $\rho$ depends on what filtering is applied to the instrument output.  After ~\cite{PhysRevD.57.4535}, with optimum filtering the maximum signal-to-noise ratio  per unit source mass $\rho'$  is 
\begin{equation} 
\label{rho-eqn}
\rho'=\sqrt{4\int_0^{\infty} W_{g_z}'(f)  df}.
\end{equation}
Prime superscripts  indicate quantities normalized by the source mass:  $\rho'=\rho/M$ and $W'_{g_z}(f)=W_{g_z}(f)/M^2.$  
It follows that the minimum detectable point mass  with $\rho=3$ is
\begin{equation}
\label{M3}
M_3=\frac 3
{\sqrt{4\int_0^{\infty}    W'_{g_z}(f) df}}.
\end{equation}
The lower panel of Figure~\ref{gradiometer-response} shows $M_3$  as a function of orbital altitude $h$ for a gradiometer limited by white spectral noise $S_g(f)=\SI{1}{mE/\sqrt{\mathrm{Hz}}}$ (approximated value for GOCE from 	\cite{touboul1999accelerometers}, \cite{{touboul1999electrostatic}}).  At  $h=\SI{330}{km},\ M_3= \SI{100}{Gt}$  and the minimum observable gradient at the peak time $t=0$ is $2\kappa_g M= \SI{0.37}{mE}.$

\section{Sensitivity of GRACE-like measurements}
\label{grace-sens}
The measurement configuration and signal parameters for the low-low satellite-to-satellite tracking (SST) of GRACE and GRACE-FO are shown in Figure~\ref{sig-parameters}, right.  The primary signal is the along-track differential position of the spacecraft, measured by microwave ranging or laser interferometry.

\subsection{Single Spacecraft Acceleration}
For simplicity, the Flat-Earth approximation  (\cite{tapley246}) is used.  This approximation neglects centrifugal acceleration, which introduces an error of less than 20\%  at Fourier frequencies greater than \SI{2}{mHz} (\cite{ghobadi2018transfer}, Figure 1; \cite{muller2017}, Figure 1.7).  In this approximation,  the acceleration on spacecraft~1 flying over point mass $M$ at along-track distance $x$ is
$a_1=-GM x/(h^2+x^2)^{3/2}.$

Define the  acceleration per unit source mass, $a'_1=a_1/M.$ Then
\begin{equation}
a'_1=-\frac{Gt v_o}{(h^2+v_o^2t^2)^{3/2}}
\end{equation}
Converting to frequency space, 
\begin{equation}
\label{a1ft}
a'_1(f) = \mathcal{F}[a'_1(t)]=\frac{4\pi i fGK_0\left(\frac{2\pi f }{f_h}\right)}{v^2_o}.
\end{equation}

\subsection{Range Acceleration Signal}
\begin{figure}[!ht]
 \center
\includegraphics[width=0.53\textwidth]{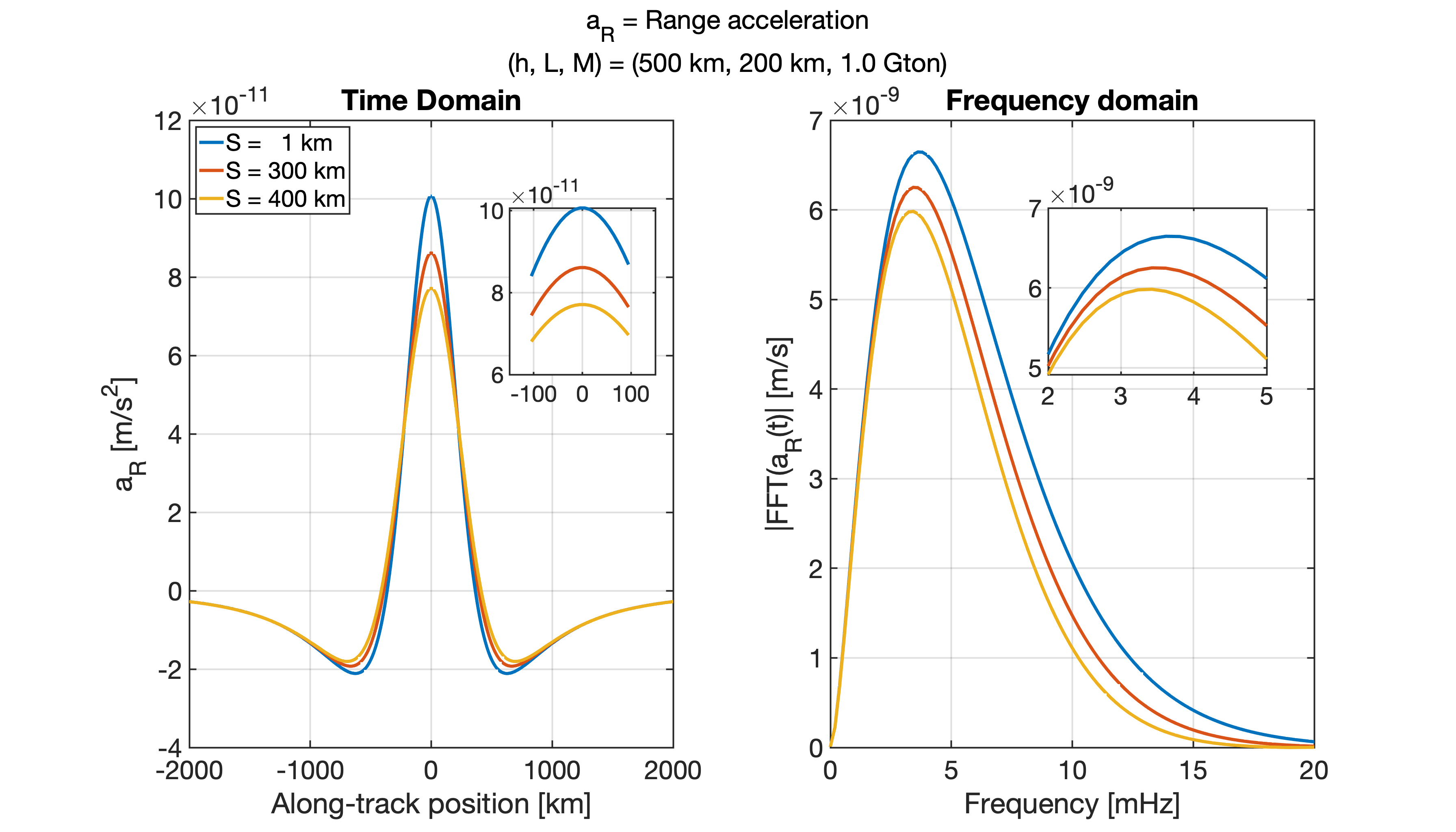}
\caption{Range acceleration resulting from a square mass centered under  the flight along-track path. Orbital altitude $h$, average spacecraft separation $L$ and source mass $M$ are indicated in the title.  The separate traces are for squares of side  S indicated in the legend.    {\it Left}:  time domain, $a_R(t)$; {\it Right}:  Fourier Transform, $a_R(f).$ \label{distr-mass}}
 \end{figure}

The acceleration experienced by spacecraft~2 is the same as spacecraft~1 at distance $L$, but delayed by $\tau =L/v_o.$  The resulting (along-track) range acceleration between  the spacecraft (Figure~\ref{distr-mass}, left) is similar to what \cite{han2013determination} computed for  the response of the GRAIL spacecraft to regional lunar gravity.   The peak range acceleration $a_R^p$ is 
\begin{equation}
a_R^p=\frac{GL}{\left(h^2+(L/ 2)^2\right)^{3/2}}M\equiv\kappa_R M,
\end{equation} 
where $\kappa_R=GL(h^2+(L/2)^2)^{-3/2}.$

Using the identity $\mathcal{F}(\text{delay } \tau)=\exp(-2\pi i f \tau)$, the  range acceleration in the frequency domain, $a_R(f),$ is given by
\begin{eqnarray}
\label{aR-of-f}
a'_R(f)&=&a'_1(f)(1-e^{-2\pi i f \tau})\\
\label{ar-abs}
|a'_R(f)|&=&2|a'_1(f)\sin(\pi f \tau)|
\end{eqnarray}
That is, in the frequency domain the range acceleration is the single-satellite acceleration multiplied by $2|\sin(\pi f \tau)|.$  For $f\ll 1/\tau,$  $|a_R(f)|~\propto L$, which is the response for the spacecraft pair acting as a gradiometer.  The response departs from that of a gradiometer at large $L,$ most conspicuously in the form of high-frequency nulls where the signal vanishes.  The first null  is at $f_\mathrm{null}=1/\tau=\SI{38}{mHz}$ for low-Earth orbit and $L=\SI{200}{km}$, as recognized by \cite{wolff1969direct}.  In  degree-variance evaluations of measurement sensitivity, the first null is expressed as a maximum in geoid height error at degree $N = f_\text{null}/f_1 = 216$ for $L=\SI{200}{km},$ and $N = 86$ for $L=\SI{500}{km},$ where $f_1 = $ orbital frequency = \SI{0.176}{mHz}.

From Equations~\ref{a1ft} and \ref{ar-abs}, 
 \begin{equation}
 \label{abs-aR}
|a'_R(f)|=\frac{8\pi fG}{v^2_o}\left|K_0\left(\frac{2\pi f}{f_h}\right)\right |
\left | \sin \left(\frac{2\pi f}{f_L}\right)\right |,
\end{equation}

where
\begin{align*}
 v_o=&\text{ orbital velocity} &= & \SI{7.6}{km/s}\\
 h = &\text{ orbital altitude} &=&   \SI{500}{km}\\
 L =& \text{ spacecraft separation} &=& \SI{200}{km}\\
  f_h=&\frac{v_o}{h}&=& \SI{15.2}{mHz}\\
  f_L=&\frac{v_o}{L/2}&=& \SI{76}{mHz}\\
  \tau=&\frac{L}{v_o}&=& \SI{26}{s}\\
  \kappa_R=&\frac{GL}{\left(h^2+(L/ 2)^2\right)^{3/2}}&=&\SI{0.101}{nm/s^2/Gt}.
\end{align*}
These numerical values apply to GRACE-FO.  
The approximately exponential attenuation with frequency of $|a_R(f)|$ has $e$-folding frequency \SI{2.5}{mHz}, corresponding to harmonic order $N=14.$ 
   
To explore the valid realm of the point-mass approximation, Figure~\ref{distr-mass} shows the range acceleration signal from a square-shaped planar mass of  side length S, computed by numerical integration.  The S=1\,km result is in agreement with the point-mass analytical calculation, which is valid at the 20\% level for sources as large as S=300\,km.  Henceforth, we restrict our analysis to the signal from a point source. 

 Equation \ref{abs-aR} gives the measurement impulse response;  that is, the range acceleration frequency response to a point mass input.   This facilitates the  direct comparison of signal and noise amplitudes as computed in the following section, and yields an expression for the minimum detectable mass for GRACE-like measurements of point source perturbations to surface gravity.  

\subsection{Noise and Mass Sensitivity}
\label{m3-section}
Consider the range measurement made by the laser ranging interferometer (LRI) on GRACE-FO.  Assuming the measurement resolution is limited by the thermal noise of the laser reference cavity (\cite{numata2004thermal}),  the displacement noise root power spectral density (rpsd) $\tilde{x}_\text{LRI}$ and strain rpsd  $\tilde{x}_\text{LRI}/L$ are given by
\begin{equation}
\label{cav-thermal}
\tilde{x}_\text{LRI}(f)/L=x_c/\sqrt{f},
\end{equation}
where $x_c$ is a constant.  For the LRI~(\cite{abich2019orbit}),  $x_c=\SI{1e-15}{}.$
The rpsd of the LRI range acceleration noise is
\begin{equation}
\label{SLRI}
\sqrt{S_\text{LRI}(f)}=(2\pi f)^2\cdot \tilde{x}_\text{LRI}(f).
\end{equation}

Take for the accelerometer measurement noise rpsd on a single satellite of GRACE and GRACE-FO (\cite{touboul1999accelerometers})
\begin{equation}
\label{accel-rpsd}
\sqrt{S_{\mathrm{ACC_1}}(f)}=\tilde{a}_0\sqrt{1+\left(\frac{f_k}{f}\right)^2}.
\end{equation}
Estaimates  of $\tilde{a}_0$ and $f_k$ range from  \SI{3e-11}{m/ s^2 /\sqrt{\text{Hz}}} and \SI{10}{mHz}, respectively (\cite{hauk2020})
to \SI{1e-10}{m/ s^2 /\sqrt{\text{Hz}}} and \SI{5}{mHz}, respectively (\cite{christophe2010orbit}, \cite{conklin2017drag}) .  We take as a compromise  $\tilde{a}_0=\SI{7e-11}{m/ s^2 /\sqrt{\text{Hz}}}$ and $f_k=\SI{5}{mHz}.$  Assuming that the acceleration measurements on the two satellites are uncorrelated, the total accelerometer noise is double:  $S_{\mathrm{ACC}}=2S_{\mathrm{ACC_1}}.$

Improved accelerometers in future missions (\cite{christophe2010orbit}, \cite{conklin2017drag}) may have $\tilde{a}_0=\SI{7e-13}{m/ s^2 /\sqrt{\text{Hz}}}.$   

\begin{figure}[!ht]
 \center
\includegraphics[width=0.52\textwidth]{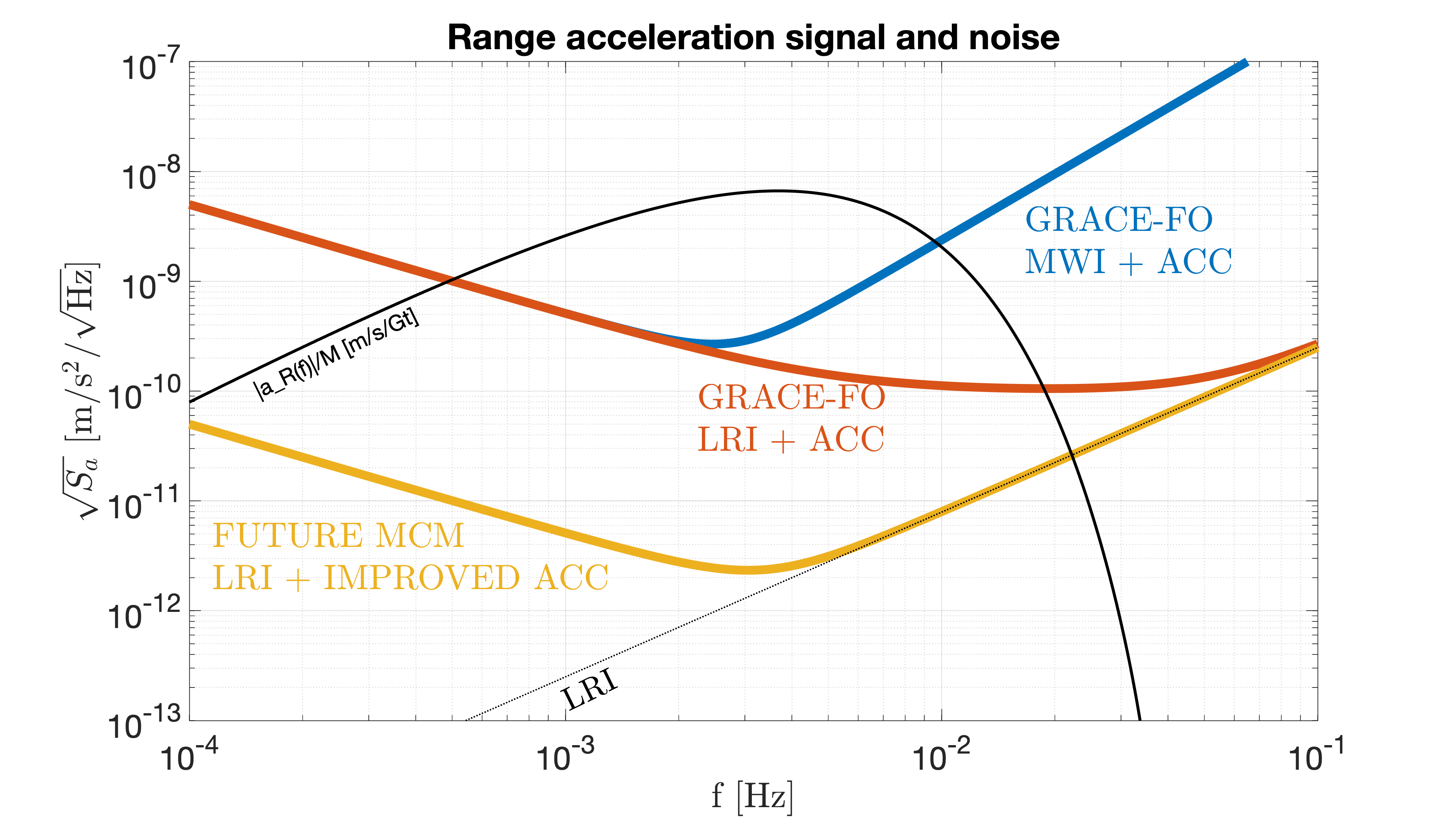}
 \caption{Total range acceleration noise rpsd $\sqrt{S_a}$ for various assumptions of instrument noise. The ranging noise for the ``GRACE-FO MWI+ACC'' is the white displacement noise of the microwave measurement on GRACE-FO, equal to \SI{6e-7}{m/\sqrt{\text{Hz}}}.  The other two noise curves assume the ranging noise of the LRI, Equation~\ref{cav-thermal}, shown as a dotted line.  Two levels of $\tilde{a}_0/[\SI{}{m/ s^2 /\sqrt{\text{Hz}}}]$ are assumed:  \SI{7e-11}{} for the GRACE-FO curves, and \SI{7e-13}{} for a future mission such as Mass Change Mission (MCM).  The solid black line is the signal spectrum  $a_R(f)$ from a \SI{1}{Gt} point mass, for  $h=\SI{500}{km}$ and  $L=\SI{200}{km},$ units m/s.  The values of $M_3$ for the three respective configurations are (Section~\ref{m3-section}) \SI{1.3}{Gt} and \SI{0.5}{Gt} and \SI{7}{Mt}. 
 \label{tot-noise-curves} }  
\end{figure}

The total instrument noise power spectral density is 
\begin{equation}
\label{Stot}
S_a=S_{\text{ACC}}+S_{\text{LRI}}.
\end{equation}
 Figure~\ref{tot-noise-curves} shows $\sqrt{S_a}$ for different ranging instrument and accelerometer noise spectra.   The MWI ranging noise is approximated by white displacement noise, $\tilde{x}_\text{MWI}=\SI{6e-7}{m/\sqrt{\rm Hz}}.$  This estimate is based on comparing MWI range measurements to simultaneous LRI range measurements.  To guide the eye to the frequencies that have the largest signal, $|a_R(f)|$ for a \SI{1}{Gt} point source from Figure~\ref{distr-mass} is overlaid as the solid black line.  The units of $|a_R|$ are m/s.
 
\begin{figure}[!ht]
\center
\includegraphics[width=0.5\textwidth]{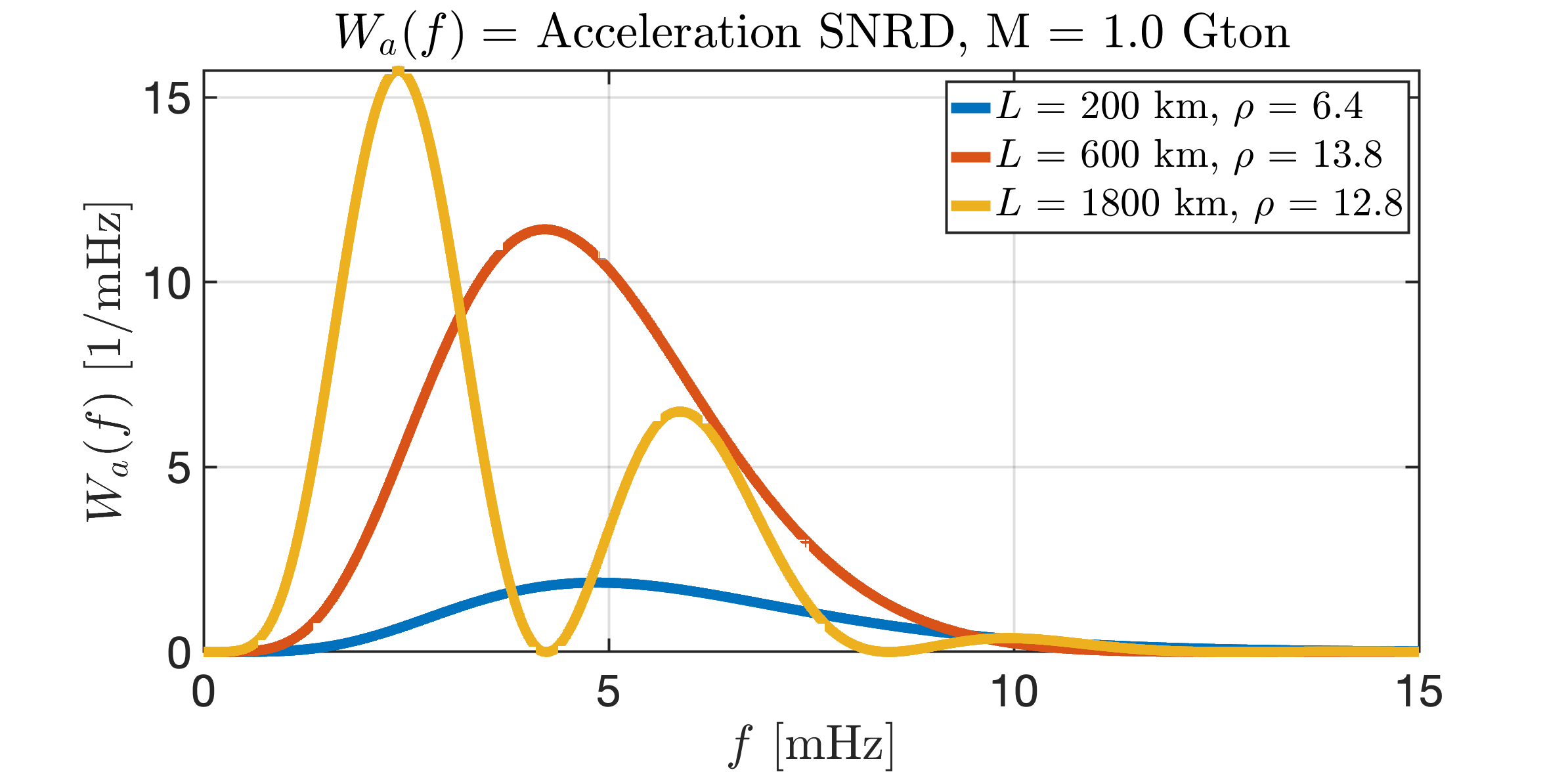}
\caption{Signal-to-noise ratio integrand of Equation~\ref{Wa-def}, $W_a(f)$ for the ``GRACE-FO LRI + ACC'' noise of Figure~\ref{tot-noise-curves}, signal from orbital  $h=\SI{500}{km}$ and source mass $M=\SI{1}{Gt}$  and three different values of spacecraft separation $L.$  The integrated signal-to-noise ratios $\rho$  from Equation~\ref{rho-eqn-ar} are indicated in the legend. \label{integrand-plot}}
\end{figure}

 As in Section \ref{grad-sec}, define the SNRD for range acceleration
 \begin{equation}
\label{Wa-def}
W'_a(f)=\frac{|a_R(f)|^2/M^2}{S_a(f)}.
\end{equation}
$W_a(f)=W_a'M^2$ is shown in Figure~\ref{integrand-plot} for several values of $L.$  The oscillations with nulls at multiples of $1/\tau = v_o/L$  degrade $\rho$ for $L$ beyond an optimum spacecraft separation.  

The optimal  signal-to-noise ratio per unit mass is
\begin{equation} 
\label{rho-eqn-ar}
\rho'=\sqrt{4\int_0^{\infty} W_a'(f)  df},\  
\end{equation}
and the source mass that gives $\rho=3$ is (cf. Equation~\ref{M3})
 \begin{equation}
\label{M3_a}
M_3=\frac 3
{\sqrt{4\int_0^{\infty}    W'_a(f) df}}.
\end{equation}

From Equations \ref{abs-aR} through \ref{Wa-def} and Equation \ref{M3_a}, the GRACE-FO parameters with the microwave ranging instrument (MWI) and LRI give respectively $M_3=\SI{1.3}{Gt}, \SI{0.5}{Gt}.$  The corresponding detectable peak accelerations, $\kappa_RM_3,$  are \SI{0.13}{nm/s^2}, \SI{0.047}{nm/s^2}.

 Another assessment  of mass sensitivity for SST laser ranging is inferred from \cite{colombo1992global}, who proposed a laser ranging mission that, with $(h, L) = (600,\ 500)\,$km was found by simulation to have  sensitivity to weekly changes of  \SI{1}{mm} water height over a square region 400\,km across,  or mass sensitivity of \SI{160}{Mt}.  In comparison, we find for the LRI on GRACE-FO at the same $(h,L)$,  $M_3=\SI{400}{Mt}.$  The two measurements have different assumed instrument sensitivity and averaging times (week-to-week vs. single-pass). 
 
\begin{figure}[!ht]
\center

\includegraphics[width=0.5\textwidth]{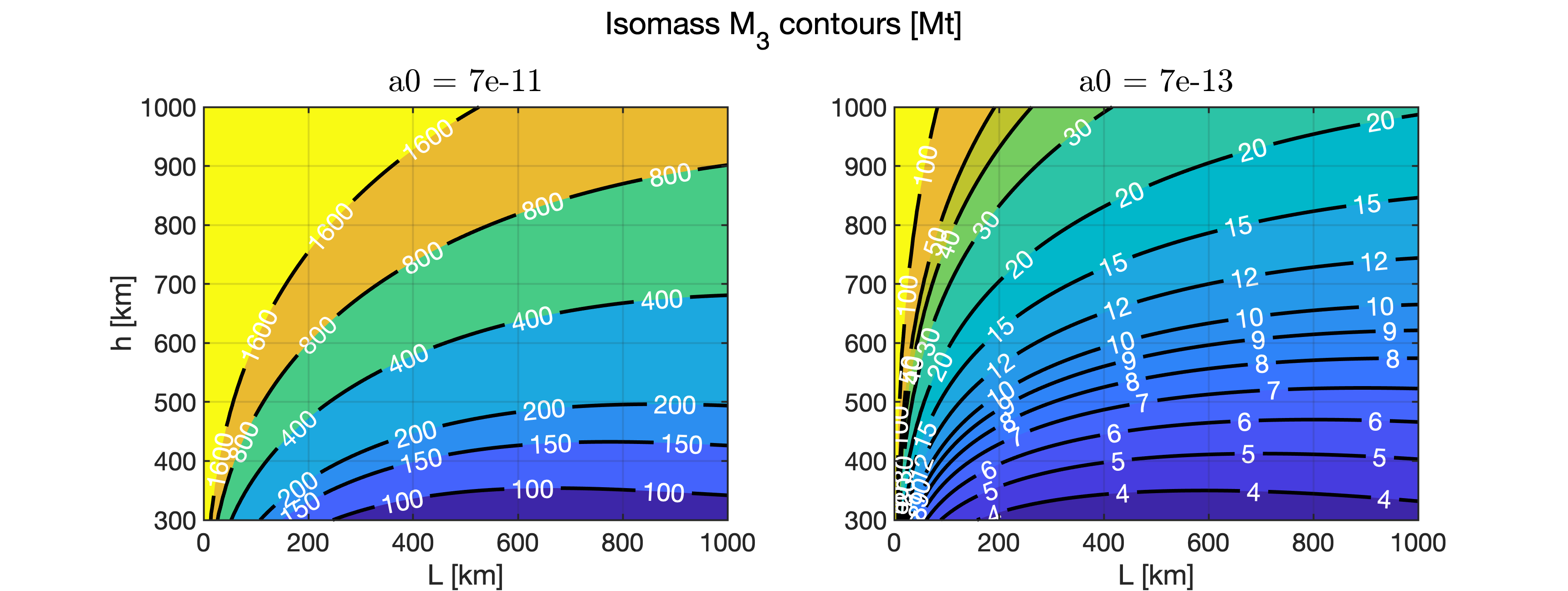}\\
\includegraphics[width=0.5\textwidth]{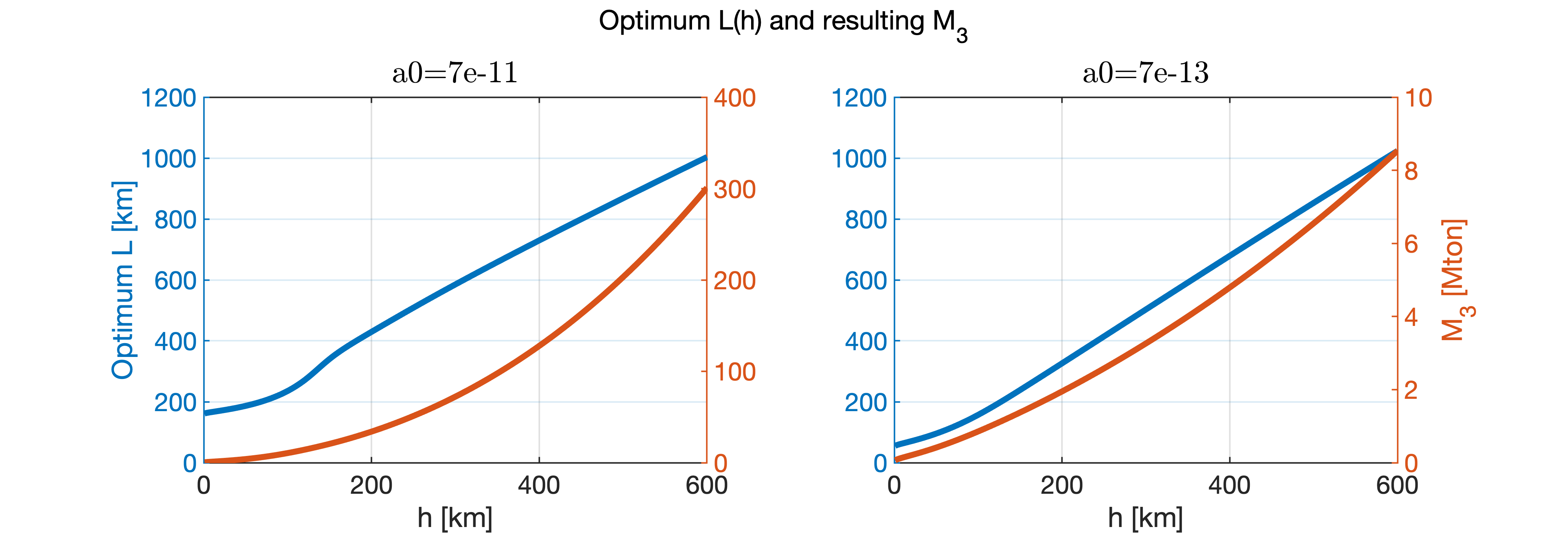}
 \caption{Mass sensitivity  of the LRI measurement on GRACE-FO {\em left}, and of a future GRACE-like mission {\em right}.  Upper row shows   isomass $M_3$ contours, in Mt, from Equation~\ref{M3_a}. \  Equation~\ref{cav-thermal} specifies the ranging noise, and accelerometer noise is given by Equation~\ref{accel-rpsd} with $\tilde{a}_0=(\SI{7e-11}{}, \SI{7e-13}{})$ \SI{}{m/ s^2/\sqrt{\text{Hz}}}, with fixed $f_k=\SI{5}{mHz}.$   Lower row shows the  optimum $L$ as a function of $h,$ (blue, left axis) and the resulting sensitivity  $M_3$  (red, right axis).   \label{contours}}
\end{figure}

Figure~\ref{contours} shows the mass sensitivity $M_3$ as a function of $h$ and $L$ for the LRI ranging instrument with two different levels of accelerometer sensitivity:   
$\tilde{a}_0=\SI{7e-11}{}$ and \SI{7e-13}{}\,\SI{}{m/s^2/\sqrt{\text{Hz}}}.    The lower row of Figure~\ref{contours} shows the optimum $L$ for a given $h$ and the resulting $M_3.$  The optimum $L$ for the LRI on GRACE-FO, operating at $h=\SI{500}{km}$, is $L=\SI{900}{km},$ which would give $M_3=\SI{200}{Mt}.$ That reflects a potential factor of 2.5 improvement over $M_3=\SI{500}{Mt}$ for the nominal satellite separation of $L=\SI{200}{km}.$  A future mission with the 
improved $\tilde{a}_0,$ $h=\SI{500}{km},$ and optimal satellite separation $L=\SI{900}{km}$ has $M_3=\SI{7}{Mt}.$
 
 \subsection{Optimal filter}
 \label{lgd-appndx}
\begin{figure}[!ht]
\center
\includegraphics[width=0.5\textwidth]{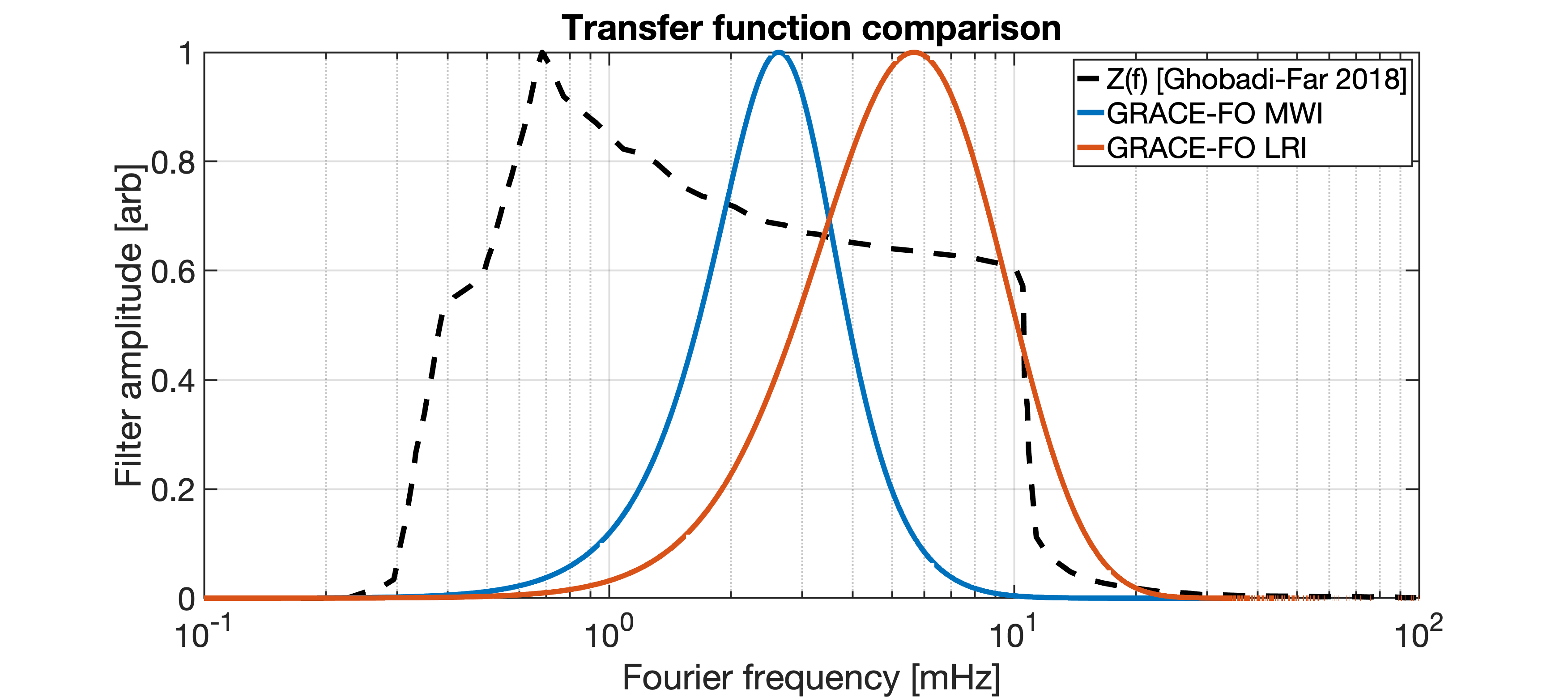}
\caption{Transfer functions for range acceleration data.  The blue trace that peaks at \SI{2.6}{mHz} is $|G(f)|$ for the MWI on GRACE-FO, and the red trace that peaks at  \SI{5.7}{mHz} is for the LRI on GRACE-FO.  $Z(f)$ is the admittance filter from \cite{ghobadi2018transfer} Figure 2(b), and applies to the MWI on GRACE-FO.   All curves  are normalized to give maximum value of 1. \label{compare-Ks}}
\end{figure}

 The filter that gives maximum signal-to-noise ratio is (\cite{wainstein1970extraction}, Chapter 3)
\begin{equation}
\label{K-of-f}
G(f)=\frac{a_R^*(f)}{S_a(f)},
\end{equation}
with~$^*$~denoting complex conjugation.  The filter's input is the measured range acceleration.  $G(f)$  is an example of a filter for extracting a signal of known waveform, in this case the range acceleration resulting from flying over a point mass.    Dropping the multiplicative constants, the filter magnitude is  
\begin{equation}
\label{Gmag}
|G(f)|=\frac{\left|f  K_0\left(\frac{2\pi f}{f_h}\right)\sin\left(\frac{2\pi f} {f_L}\right)\right|}{S_a(f)}.
\end{equation}
Normalized $|G(f)|$ for the MWI and  LRI on GRACE-FO are shown  in Figure~\ref{compare-Ks}.

\cite{ghobadi2018transfer}  analyzed the GRACE-FO MWI signal in terms of the line-of-sight gravity difference, and applied the same analysis method to the GRACE-FO LRI signal in \cite{ghobadi2020}.  In their analysis of MWI data, \cite{ghobadi2018transfer} defined the gravimetric quantity $\delta g^\text{LOS}_{12},$ or line-of-sight (LOS) gravity difference, which differs from the range acceleration residual $\delta \ddot{\rho}$ by $\Delta_0,$ the residual centrifugal acceleration:
\begin{equation}
\delta g_{12}^\text{LOS}=\delta \ddot{\rho}+\Delta_0.
\end{equation}
Residuals are relative to a reference field.
The admittance $Z(f)$ is defined as the ratio of power spectra,
\begin{equation}
\label{Zdef}
Z(f)=\frac{S_{\delta\ddot{\rho},\delta g_{12}^\text{LOS}}(f)}
{S_{\delta\ddot{\rho},\delta\ddot{\rho}}(f)},
\end{equation}
where $S_{\delta\ddot{\rho},\delta\ddot{\rho}}(f)$ is the power spectrum of the MWI range acceleration measurement and $S_{\delta\ddot{\rho},\delta g_{12}^\text{LOS}}(f)$ is the cross-power spectrum between the range acceleration and the LOS gravity difference. Keeping the shorthand notation $p(f)={\cal F}[p(t)],\ Z(f)$ is a filter that transforms residual range acceleration $\delta\ddot{\rho}(f)=a_R(f)$ to $\delta g_\text F^\text{LOS}(f)$, an estimate of $\delta g^\text{LOS}_{12}(f):$
\begin{equation}
\delta g^\text{LOS}_\text F(f)=Z(f)\delta\ddot{\rho}(f).
\end{equation}

 $Z(f),$ normalized to have a maximum value of 1, is shown as the dashed trace in Figure~\ref{compare-Ks}.   $Z(f)$ is the optimal filter to apply to MWI range acceleration,  based on the measurement data that includes signal from the gravity field.  It applies to extracting the best SNR from a residual regional or global field and does not explicitly depend on instrument noise spectra.

In contrast, $G(f)$ is fine-tuned to the problem of detecting the specific waveform of a point mass, in the presence of known measurement noise.   Since  a point mass generates a field with the highest possible frequency content, the $G(f)$ passband starts higher in frequency than $Z(f).$  The LRI $G(f)$ passband is higher than for the MWI because the LRI measurement has reduced noise at high frequency.    

A practical use for the $G(f)$ filter is searching for unknown point-like features, such as underground water storage of 100\,km spatial extent.  The filter would be applied to range acceleration measurements after subtracting the effect of the known field, including time-varying gravity, and non-gravitational accelerations. 

 \section{Conclusion}
 We derived the optimum sensitivity of orbiting gravimetric satellites to a point source, that is a single mascon.  The signal-to-noise ratio is found as a function of instrument noise and orbital parameters.  The signal is converted to frequency space by the Fourier transform, and the signal-to-noise ratio is derived from optimal filtering a signal of known waveform.  This analysis differs from the conventional approach of spherical harmonic expansion to characterize the field from an arbitrary mass distribution.  Such an expansion requires a very large harmonic order to accurately approximate the field from a point source, as shown   in~\ref{appa}.
  
The frequency response of an orbiting gradiometer to a point mass directly under the flight track is approximated by Equation \ref{FT-gz} that depends only on the orbital altitude and the magnitude of the point mass. Likewise, for an SST-based measurement of the gravitational field,  the range acceleration is approximated by Equation \ref{abs-aR} that includes dependency on the average satellite separation.  Applying Wiener optimal filter theory, these responses and the noise spectra of the ranging measurement and of accelerometer-based measurement of non-gravitational forces give  $\rho,$ the  maximum achievable signal-to-noise ratio.  The resolvable  mass $M_3$ is defined as the magnitude of the point mass that gives $\rho=3.$  $M_3$ is the ultimate  mass sensitivity, and realistic non-point mass distributions that are not directly under the flight track will give larger $M_3$ in practice.  Nonetheless, $M_3$ provides a figure of merit for comparing future missions with different orbits and instrument sensitivities to  guide the design of such missions.  For SST measurements $M_3$ has a minimum value at a calculable satellite separation $L,$ giving the optimum separation for discovering point-like (meaning less than approximately 300\,km) features such as subsurface water storage.  Equation~\ref{Gmag} specifies the optimal  filter for such a search.  As a caveat, the $M_3$ metric and its $L$ optimization does not apply to large-scale gravimetry, such as required by oceanography.
 \section*{Acknowledgments}

The author thanks  Kirk McKenzie,  Gabriel Ramirez, Pep Sanjuan and David Wiese for useful discussions, and Christopher McCullough for key insights.   The contributions of four anonymous reviewers, who suggested improvements that are incorporated in this manuscript, is gratefully acknowledged.  This research was carried out at the Jet Propulsion Laboratory, California Institute of Technology, under a contract with the National Aeronautics and Space Administration. \textcopyright 2020 California Institute of Technology. Government sponsorship acknowledged.

 \appendix
 \section{Multipole expansion and the Wahr equation for surface density}
 \label{appa}
\begin{figure}
\begin{center}
\includegraphics[width=0.4\textwidth]{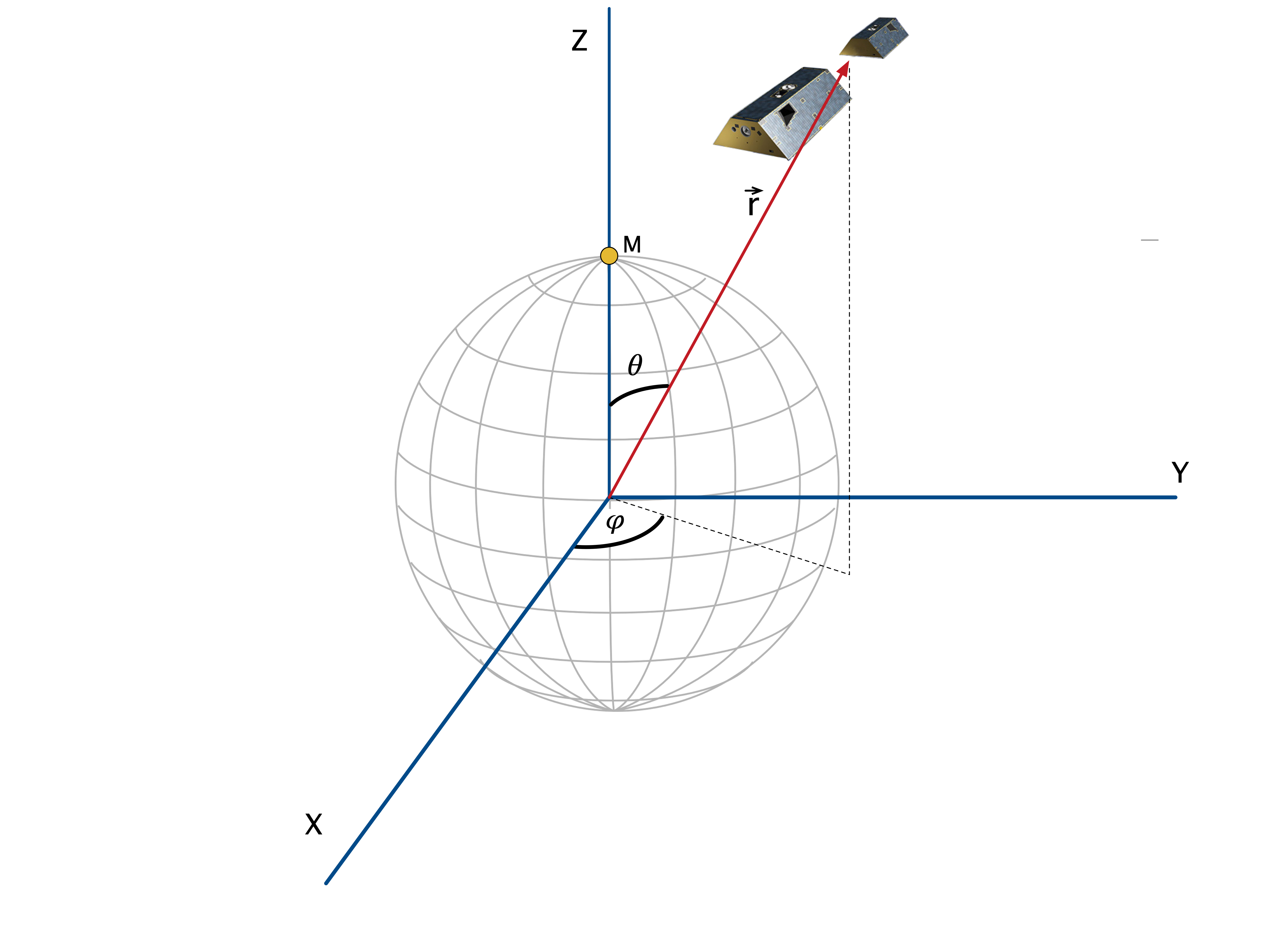}
\caption{Coordinate system for the spherical harmonic expansion of geopotential, \cite{kaula2013theory}.  The satellite constellation position, defined as the center of mass for a gradiometer or (illustrated) the center of the line of sight between two SST satellites relative to the center of the Earth, is $\vec{r}(r,\theta,\phi)$, where $r$ is the distance from the center of the Earth, $\theta$ is the co-latitude, and $\phi$ is the longitude.\label{earth-sphere}}
\end{center}
\end{figure}

The gravitational potential  is conventionally expressed as the multipole expansion  (\cite{kaula1966tests},  \cite{kaula2013theory}, \cite{chao1987changes})
        \begin{align}
        \label{kaula-potential}
        U(r,\theta,\phi)=&\frac{GM_e}{a}\Dsumz\left(\frac a r\right)^{n+1}\Pnm(\cos \theta)\\ \nonumber
                                 &\times(\Cnm\cos m\phi+\Snm\sin m\phi).
        \end{align}
As illustrated in Figure \ref{earth-sphere}        $(r, \theta, \phi)$ = (distance from the center of the Earth, co-latitude, longitude),        
 $(a, M_e) =$  (Earth radius, Earth mass), and $\Pnm$ is the fully-normalized associated Legendre function. The field is entirely specified by the Stokes coefficients $(\Cnm, \Snm).$ 
 
  For a known mass distribution $dM = \rho(r',\theta',\phi')dV'$ with primes designating  the source mass coordinates, $(\Cnm, \Snm)$  are evaluated as the volume integral
 \cite{L2handbook2018} 
         \begin{equation}
        \label{solve-for-C-S}
        \begin{bmatrix} \Cnm \\ \Snm \end{bmatrix}
         =\frac{1}{(2n+1)M_e}\int_{V'} dM\,\left(\frac{r'}{a}\right)^n\Pnm(\lambda')
         \begin{bmatrix}
         \cos m \phi' \\ \sin m \phi'
         \end{bmatrix},
        \end{equation}
where \ $\lambda'=\cos\theta'.$ For a point mass at $r'=a$ 
\begin{equation}
\label{C-S-for-point-mass}
 \begin{bmatrix}
         \Cnm \\ \Snm
                \end{bmatrix}
=\frac{M}{(2n+1)M_e}\Pnm(\lambda')
         \begin{bmatrix}
         \cos m \phi' \\ \sin m \phi'
         \end{bmatrix}.
        \end{equation}
 A single point mass can be placed at the north pole, $(\theta',\phi')=(0,0)$ without loss of generality.  Then
 \begin{equation}
 \label{C-S-point-mass-at-pole}
 \begin{bmatrix}
         \Cnm \\ \Snm
                \end{bmatrix}
=\frac{M}{(2n+1)M_e}\Pnm(1)
         \begin{bmatrix}
         1 \\ 0
         \end{bmatrix}.
        \end{equation}
The relationship between the fully normalized Legendre function $\overline{P}_{nm}$ and the associated Legendre function $P_{nm}$ is               
\begin{equation}
\Pnm=\sqrt{\frac{(2-\delta_{m0})(2n+1)(n-m)!}{(n+m)!}}P_{nm},
\end{equation}
where $\delta_{mn}$ is the Kronecker delta.
Since $P_{nm}(1)=\delta_{m0},\ 
\Pnm(1)=\delta_{m0}\sqrt{2n+1}.$
For a point mass at the pole Equation~\ref{C-S-for-point-mass}  reduces to
\begin{equation}
 \label{C-S-point-mass-at-pole-final}
 \begin{bmatrix}
         \Cnm \\ \Snm
                \end{bmatrix}
=  \frac{M}{M_e}\frac{1}{\sqrt{2n+1}} \begin{bmatrix} \delta_{m0}\\ 0          
         \end{bmatrix}.
        \end{equation}
The potential from Equations \ref{kaula-potential} and \ref{C-S-point-mass-at-pole-final} is independent of $\phi$ and is given by the multipole expansion
\begin{equation}
\label{U-point}
U(r,\theta)=\frac{GM} {a} \sum_{n=0}^{\infty}\left(\frac {a} r \right)^{n+1}P_n(\cos\theta),
\end{equation}
the familiar expansion from electrostatics  for the azimuthally symmetric electric field from a point charge (\cite{jackson2007classical}) and from  gravitational potential theory (\cite{blakely1996}, Section 6.4.2).

The Wahr equation for surface density from  $(\Cnm,\Snm)$ (\cite{wahr1998time}) is
 \begin{align}
 \label{wahr-water-height}
 \nonumber
 \sigma(\theta',\phi')=&\frac {a\rho_\text{ave}}{3}\Dsumz \Pnm(\cos\theta')\frac{2n+1}{1+k_n}\\
 &\times(\Cnm\cos m\phi'+\Snm\sin m\phi'),
  \end{align}
  where $k_n=$ Love number.

To study the error of a finite-degree spherical harmonic approximation to a point mass, consider a spherical cap in the limit of small cap size.   
 The spherical cap is centered at  coordinates $(\theta',\phi')$  and its angular radius is $\alpha$ and $\lambda \equiv \cos\alpha.$ As computed by \cite{pollack1973spherical}, the Stokes coefficients are 
  \begin{equation}
 \label{cap-anywhere}
 \begin{bmatrix}
 \Cnm\\ \Snm
 \end{bmatrix} = -\frac{M}{M_e}\frac{P_{n+1}-P_{n-1}}{(2n+1)^2(1-\lambda)} \Pnm(\cos\theta')
 \begin{bmatrix}
 \cos m\phi' \\ \sin m \phi'
\end{bmatrix},
 \end{equation} 
where we use the shorthand $P_j(\lambda)=P_j.$ For a spherical cap at the north pole,
 \begin{equation}
 \label{cap-pole}
 \begin{bmatrix}
 \Cnm\\ \Snm
 \end{bmatrix} 
 = -\frac{M}{M_e}\frac{P_{n+1} -P_{n-1} }{(2n+1)^{3/2}(1-\lambda)} 
   \begin{bmatrix} \delta_{m0}\\ 0         
 \end{bmatrix}.
 \end{equation} 
The spherical cap reduces to a point mass in the limit of $\alpha=0, \text{ or }\lambda =1$;  substituting 

$\lim_{\lambda \to 1}\left[P_{n+1} -P_{n-1} \right]=(\lambda-1)(2n+1)$
= into Equation \ref{cap-pole} gives Equation~\ref{C-S-point-mass-at-pole-final}.

By comparing expressions similar to Equation \ref{kaula-potential} and Equation \ref{wahr-water-height}, \cite{dickey1997satellite} identifies
\begin{equation}
\hat{\overline {C}}_{nm}+\hat{\overline {S}}_{nm}=\frac{\rho_\text{ave}}{3 \rho_w}\frac{2n+1}{1+k_n}(\Cnm+\Snm)
\end{equation}
where $\rho_w$ is the density of water as the transformation to convert geoid expansion coefficients $(\overline {C}_{nm}, \overline {S}_{nm})$ to  mass expansion coefficients $(\hat{\overline {C}}_{nm},\hat{\overline {S}}_{nm})$,  p. 101 their Equation~(B5).

At the pole, from Equation  \ref{wahr-water-height}, dropping the $n=0$ term that represents the total potential of the Earth, and neglecting the Earth's elasticity by setting $k_n=0,$
\begin{equation}
\sigma= \frac{a\rho_\text{ave}}{3}\sum_{n=1}^{\infty}\Pnm(1)(2n+1)\Cnm.
\end{equation}
From Equation~\ref{cap-pole},
\begin{eqnarray}
\label{sigma-inf-sum}
\nonumber \sigma &=&-\frac{M}{M_e}\frac a {1-\lambda}  \frac{\rho_\text{ave}}{3}\\ 
\nonumber&&\times\sum_{n=1}^{\infty}\overline{P}_{nm}(1)\frac{2n+1}{(2n+1)^{3/2}}(P_{n+1} -P_{n-1} )\\
\nonumber&=&-\frac{M}{M_e}\frac a {1-\lambda}  \frac{\rho_\text{ave}}{3}\sum_{n=1}^{\infty}P_{n+1} -P_{n-1} \\
\label{sigma-full-sum}
&=&\frac{M}{M_e}  \frac{a\rho_\text{ave}}{3}T_{\infty}.
\end{eqnarray}

\begin{figure}[!ht]
\begin{center}
\includegraphics[width=0.5\textwidth]{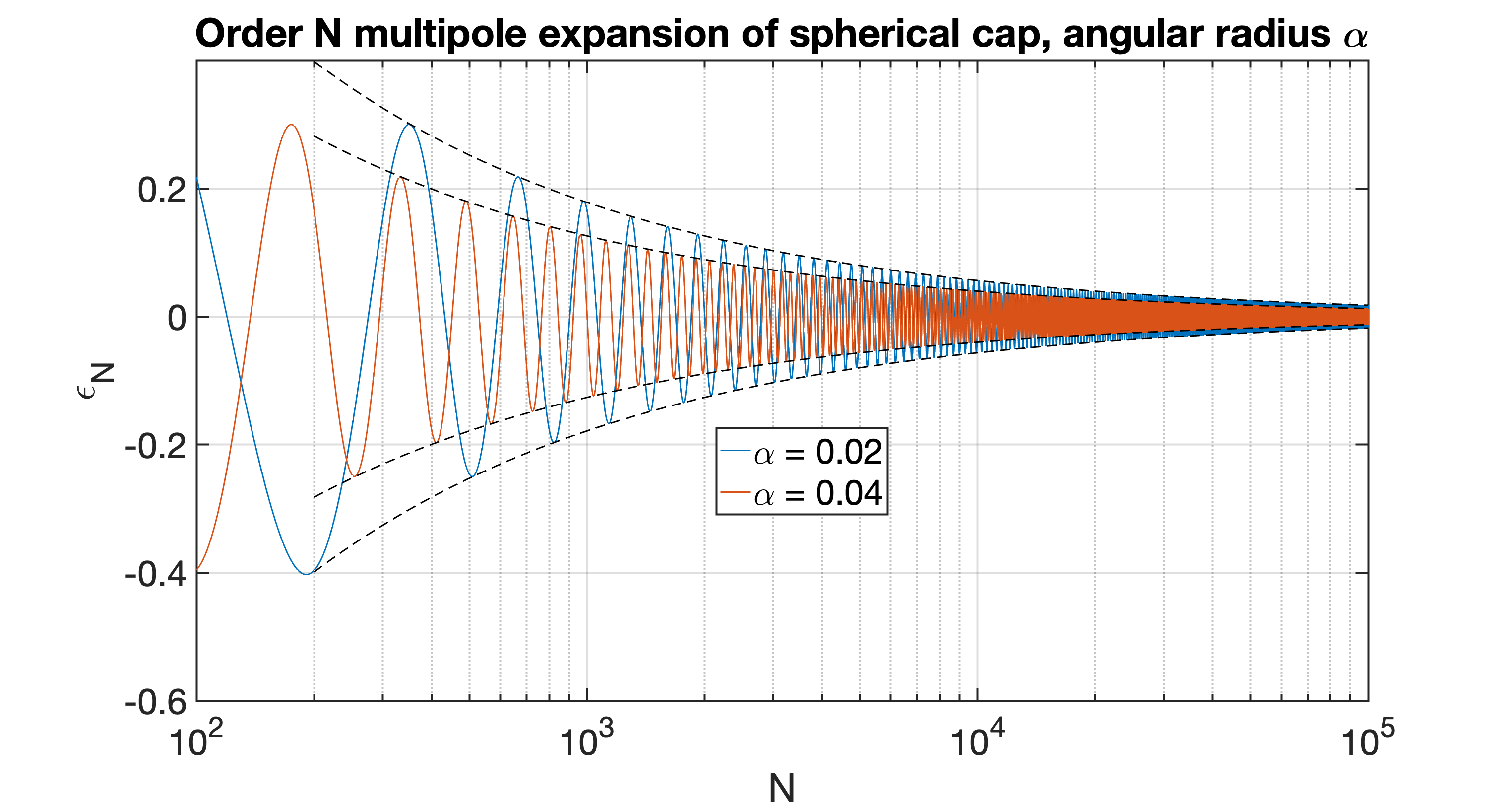}
\caption{Truncation error, Equation \ref{epsilon-N}, in representing the field from a  mass of small spatial extent by spherical harmonic expansion of order $N.$  The dashed lines follow  the large $N$ asymptote envelope, $\pm P_N(\cos\alpha)\sim \pm\sqrt{2/(\pi N \sin\alpha)}.$ \label{trunc-err}}
\end{center}
\end{figure}

The quantity $T_{\infty}$ is the $N=\infty$ limit of  the truncated sum, defined as
\begin{eqnarray}
\nonumber T_N(\lambda)&=& \frac 1{1-\lambda}     \sum_{n=1}^N  P_{n-1}-P_{n+1} \\
\nonumber &=& \frac 1{1-\lambda}    \left( P_0+P_1 -(P_N+P_{N+1}\right)\\
&=&\frac{1+\lambda-(P_N+P_{N+1})}{1-\lambda}.
\end{eqnarray}
For a small spherical cap, $\alpha <<1$ (and $\lambda = \cos \alpha$ slightly $<1$), the cap area is $A=\pi (a\alpha)^2.$  Using $\sigma=M/A \text{ and }\rho_\text{ave}=3M_e/(4\pi a^3),$
 Equation \ref{sigma-full-sum} is equivalent to
  \begin{equation}
 \label{Tinf}
 \frac{\alpha^2 T_{\infty}(\cos\alpha)} 4-1=0.
 \end{equation}
The fractional error in $\sigma$ due to truncation of the summation Equation~\ref{sigma-inf-sum} at order  $N$ is
 \begin{equation}
 \label{epsilon-N}
  \epsilon_N= \frac{\alpha^2 T_N(\cos\alpha)} 4-1\approx -P_N(\cos\alpha).
\end{equation}
See Figure~\ref{trunc-err} for $\epsilon_N$ with small spherical caps of two different sizes.  The slow reduction of $|\epsilon_N|$ with increasing $N$  shows that  the unfiltered spherical harmonic expansion is ill-suited to characterize the field from a point-like source.  The truncation error is often  reduced by applying a spectral localizing filter (\cite{panet2013earth}, Appendix 2);  see also \cite{wahr1998time},  \cite{swenson2002methods}, \cite{seo2005filters}, and \cite{werth2009evaluation}.

\bibliographystyle{apalike}
\bibliography{all-all-grace}

\end{document}